# EXTENSION OF THE C STAR ROTATION CURVE OF THE MILKY WAY TO 24 KPC

P. BATTINELLI[1], S. DEMERS[2*], C. ROSSI[3**], K. S. GIGOYAN[4]

[1] INAF, Osservatorio Astronomico di Roma, Viale del Parco Mellini 84, 00136 Roma, Italy
E – mail: paolo.battinelli@esteri.it

[2] Département de Physique, Université de Montréal, CP 6128, Succursale Centre-Ville, Montréal, Qc, H3C 3J7, Canada ( [*] Guest investigator, Dominion Astrophysical Observatory, Herzberg Institute of Astrophysics, National Research Council of Canada )
E – mail: demers@astro.umontreal.ca

[3] Dipartimento di Fisica, Universita "La Sapienza", Piazzale a. Moro 2, 00185 Roma, Italy )
( [**] Guest investigator at the Asiago Astronomical Observatory, INAF, Italy )
E-mail: corinne.rossi@uniroma1.it

[4] V. A. Ambartsumian Byurakan Astrophysical Observatory, 0213, Aragatsotn Marz, Armenia
E-mail: kgigoyan@bao.sci.am

## ABSTRACT

Demers and Battinelli published, in 2007 the rotation curve of the Milky Way based on the radial velocity of carbon ( C ) stars outside the Solar circle. Since then we have established a new list of candidates for spectroscopy. The goal of this paper is to determine the rotation curve of the Galaxy, as far as possible from the Galactic center, using N – type C stars. The stars were selec - ted from their dereddened 2MASS colours, then the spectra were obtained with the Dominion

Astrophysical Observatory and Asiago 1.8 m telescopes. This publication adds radial velocities and Galactrocentric distances of 36 C stars, from which 20 are new confirmed. The new results for stars up to 25 kpc from the Galactic center, suggest that the rotation curve shows a slight decline beyond the Solar circle.

# РАСШИРЕНИЕ КРИВОЙ ВРАЩЕНИЯ УГЛЕРОДНЫХ ЗВЕЗД МЛЕЧНОГО ПУТИ ДО 24 КПК

## П. БАТТИНЕЛЛИ[1], С. ДЕМЕРС[2], К. РОССИ[3], К. С. ГИГОЯН[4]

Демерс и Баттинелли издали, в 2007 кривая вращения Млечного Пути, основанного на радиальной скорости углеродных ( C ) звезд вне Солнечного круга. С тех пор мы установили новый список кандидатов для спектроскопии. Целъ данной работы является определением кривую вращения Галактики, в максимально возможной расстоянии от Галактического центра, использованием C звезд класса N. Звезды отобраны, учитывая покраснение в их 2MASS цветах, а спектры были получены на 1.8 м телескопах обсерватории Доминиона и Азиаго. Эта публикация добавляет радиальные скорости и Галактоцентрическое расстоянии 36 C звезд, из которых 20 подтверждены впервые. Новые результаты, для звезд до 25 кпс расстоянии от центра Галактики, предлагают, что кривое вращение показывает небольшой уклон вне Солнечного круга.

*1. Introduction.* The surface mass distribution of the Milky Way is still poorly constrained, even while our knowledge of Galactic stellar abundance distribution grows ever more detailed. One reason for this shortcoming is the difficulty to determine the Galactic rotation curve in the outer disk. Battinelli and Demers [ 1 ] discovered numerous carbon ( C ) stars in the outer disk of M31 reaching 40 kpc. Such C stars are certainly also present in the Milky Way. Indeed, N – type C stars can be used as kinematical probes. They are intrinsically bright, $< M( I ) > = - 4.6$ [ 2 ]. Moreover, these intermediate – age stars are old enough to have lost memories of their initial kinematic conditions with which they might have been ( i. e. motion within parent cluster / association ). At the same time, these stars are still relatively young, with smaller random velo – cities than older tracers ( e. g. planetary nebulae ). They are believed to be members of the thin disk population [ 3 ].

An important characteristics of the N – type C stars is their small dispersion in the absolute magnitude, which allows to use as reliable candles when some color criteria are adopted ( see papers [ 4, 5 ], for more details ).

Demers and Battinelli [ 6 ] established the Galactic rotation velocity up to distance of 15 kpc using genuine N – type C stars. In that paper they adopted the following common criteria for the intrinsic colors to select disc candidates, $( J - K )_o > 1.4$ and $( H - K )_o > 0.45$, plus additional restrictions on photometric limits to avoid high reddening regions and dust – enshrouded C stars which would lead to unreliable results.

The present paper is a follow – up of the publication [ 6 ]. We refer to paper [ 6 ] for a detailed description of the methods for candidate selection, data reduction, radial velocity determination, distance estimates. Radial velocity targets are selected from the 2MASS ( Two Micron All – Sky Survey Catalogue ) [ 7 ] with colors corresponding to N – type C stars. Stars are de – reddened using Schlegel et al. [ 8 ] reddening software. We exclude stars within $3^o$ from the Galactic plane to avoid high uncertain reddening. The new data set includes targets that are calculated to be as far as 24 kpc from the Galactic center. To reach that far we had to relax our selection criteria to include candidates with Galactic latitiudes $3° < |b| < 9°$. We then become limited by faintness of the candidates which cannot be properly acquired by the guiding system of the telescope.

*2. New Observations.* The new spectra, discussed in this paper, were obtained between July 2007 and January 2010. We used the spectrograph attached to the Cassegrain focus of the Dominion Astrophysical Observatory ( DAO, Canada ) 1. 8 m Plaskett Telescope. The same spectroscopic setup for the observations in October 2006 was selected. The spectral region cove - red ranges from 6100 Å to 6800 Å. Because our previous experience showed that 24 % of the selected targets on the basis of $(J - K)_o$, are not carbon stars but mostly emission – line objects. In 2009 we initiated a low – dispersion spectroscopic survey, with the 1.8 m telescope at Asiago ( Italy ), equipped with the Asiago Faint Object Spectrometer and Camera( AFOSC ), to confirm the nature of the stars to be observed at the DAO. We obtained spectra in the range 3700 – 8000 Å ( grism No 4, dispersion 4.2 Å/pixel ).

Further more, in addition to those used during the previous runs, we selected in 2007 and 2010 three other carbon stars for radial velocity standards. They are CGCS 4285 ( CGCS – Cool Galactic Carbon Stars, Alksnis et al.)[ 9 ] )( $V_h = +50$ km/s ) and CGCS 10 ( $V_h = -70.2$ km/s ). Their radial velocities are published by Aaronson et al. [ 10 ] ) and Metzger and Schechter [ 11 ]. To avoid large uncertainties in the rotation velocity, no C stars closer than 45° from the anti - center were observed. This survey is still active and being out in collaboration with Byurakan Astrophysical Observatory ( BAO, Armenia ).

In Figure 1a and 1b we present a few examples of spectra of new confirmed C stars.

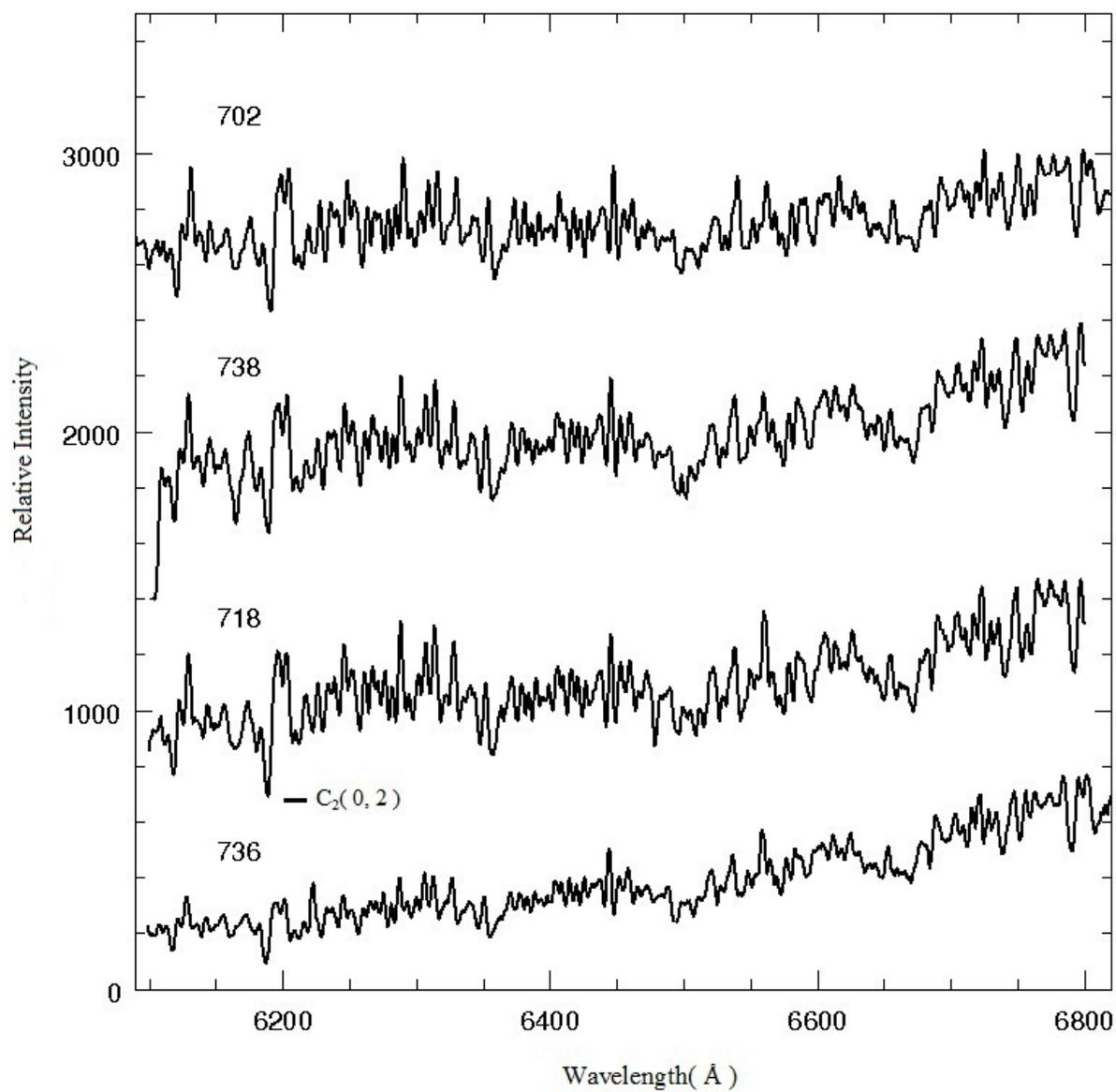

Figure 1(a). Examples of higher - resolution DAO spectrum( Fig. 1a) of the new C stars in the range 6100 - 6800 Å. Spectra are smoothed by 5 pixel box car. The absorption band of the $C_2$ ( 0, 2 ) Swan system at bandhead 6192 Å is indicated.

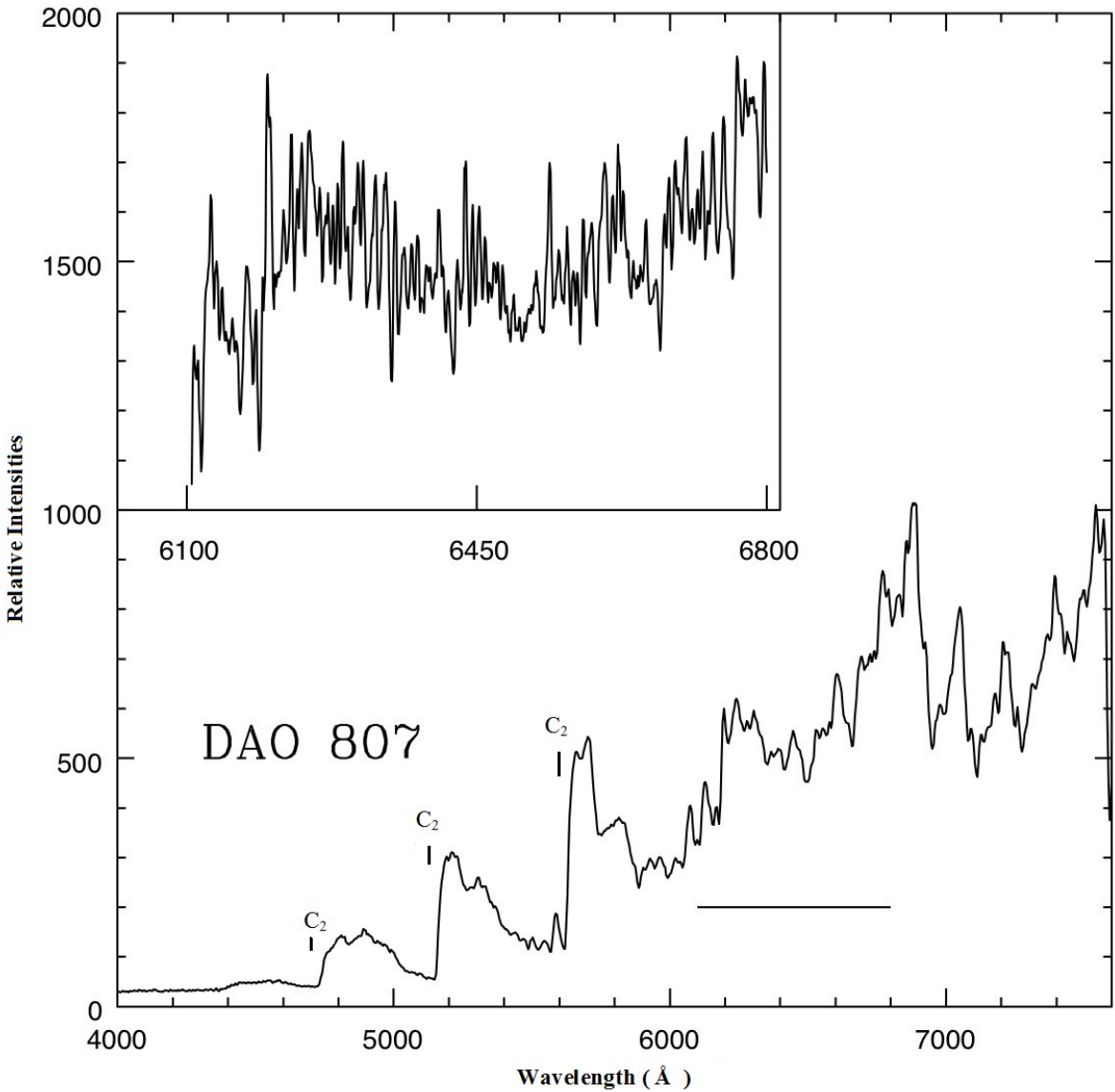

Figure 1(b). Example of a low – resolution spectrum obtained at Asiago, on January 4, 2009. Swan bands of the $C_2$ molecule are indicated. The horizontal line indicate the spectral range of the DAO spectrum, shown in the inset.

*3 . Data Analysis.* This publication adds 36 new radial velocities and Galactocentric distances of spectroscopically confirmed C stars. The stars are listed in order of their J2000.0 right ascen – sion in Table 1. We give our own running numbers because many of them are not in the CGCS Catalogue [ 9 ], *l* and *b* are Galactic coordinates, $d_G$ is the distance to the Galactic center in kpc and the Heliocentric velocities with errors are in km/s., when available, the association with the CGCS [ 9 ], is also reported in the Table 1.

We discovered, after the spectra were taken, that star No 718 is among the stars observed by Aaronson et al. [ 10 ]. They quote $V_h$ = - 72 km/s while we have $V_h$ = - 69±2km/s. The Heliocentric distance of each star is estimated using its apparent K magnitude, reddening and absolute K magnitude based on its ( J – K )$_o$ and following the relation between $M_K$ and the intrinsic ( J – K )$_o$ color, used in Demers and Battinelli [ 6 ].

$$M_K = - 6.31 – 0.99( J – K )_o \qquad (1)$$

Table 1  Data For 36 Spectroscopically Confirmed Objects

| Star No | RA. ( J2000 ) DEC. | *l* | *b* | $d_G$( kpc) | $V_H$( km/s ) | CGCS Association [ 8 ] |
|---|---|---|---|---|---|---|
| 701 | 19$^h$ 25$^m$ 57.7$^s$ +25° 01' 36" | 59°.0 | +4°.1 | 13.8 | -127 ±3 | |
| 702 | 19 27  41.6  +24 46 59 | 50.0 | +3.7 | 15.7 | -108±1 | |
| 703 | 19 31  18.9  +26 48 16 | 61.2 | +3.9 | 12.7 | -113±4 | |
| 704 | 19 36  11.3  +27 15 47 | 62.1 | +3.2 | 12.1 | -115±5 | |
| 705 | 19 38  39.4  +28 37 01 | 63.5 | +3.4 | 17.1 | -96±2 | |
| 706 | 19 39  33.9  +15 50 23 | 52.5 | -3.1 | 12.2 | -5±3 | |
| 707 | 19 40  07.8  +15 33 23 | 52.3 | -3.4 | 15.5 | -91±4 | |
| 708 | 19 41  14.1  +30 26 14 | 65.4 | +3.8 | 14.2 | -108±5 | |
| 709 | 19 41 45.8  +29 09 54 | 64.3 | +3.0 | 13.0 | -101±1 | |
| 710 | 19 42  35.3  +29 53 19 | 65.1 | +3.2 | 12.7 | -92±8 | |
| 712 | 19 44 45.5  +17 12 54 | 54.7 | -3.3 | 17.7 | -52±7 | |
| 714 | 19 49 40.5  +34 21 57 | 69.7 | +4.2 | 13.2 | -125±6 | CGCS 4510 |
| 715 | 19 52 53.2  +20 51 09 | 58.4 | -3.3 | 15.2 | -113±1 | CGCS 4539 |
| 717 | 19 58 58.2  +36 13 38 | 72.3 | +3.5 | 12.3 | -100±4 | CGCS 4592 |
| 718 | 20 02 14.0  +36 42 02 | 73.0 | +3.2 | 13.4 | -69±2 | CGCS 4625 |
| 719 | 20 02 56.7  +37 00 56 | 73.4 | +3.2 | 14.8 | -135±4 | |
| 721 | 20 08 55.9  +39 28 35 | 76.1 | +3.6 | 14.9 | -144±2 | |
| 722 | 20 14 48.9  +40 12 17 | 77.3 | +3.0 | 15.2 | -130±1 | CGCS 4773 |
| 725 | 20 19 53.7  +41 47 29 | 79.2 | +3.1 | 13.7 | -48±11 | |
| 726 | 20 27 39.8  +31 12 20 | 71.4 | -4.2 | 15.2 | -137±3 | |
| 845 | 21 30 38.5  +45 06 55 | 90.0 | -4.5 | 18.1 | -120±2 | CGCS 5330 |
| 1007 | 20 52 32.1  +30 34 01 | 74.1 | -8.9 | 16.9 | -117±4 | |
| 728 | 20 54 31.5  +39 42 11 | 81.5 | -3.4 | 13.8 | -126±3 | CGCS 5030 |
| 1008 | 21 09 28.0  +34 51 07 | 79.8 | -8.8 | 24.0 | -173±5 | |
| 1010 | 21 59 06.0  +45 39 01 | 94.2 | -7.4 | 17.6 | -165±6 | |
| 732 | 22 06 19.1  +61 00 53 | 104.2 | +4.3 | 13.4 | -71±2 | CGCS 5580 |
| 1011 | 22 20 57.5  +47 23 23 | 98.3 | -8.1 | 21.3 | -120±5 | CGCS 5633 |
| 733 | 22 51 10.4  +64 14 57 | 110.3 | +4.4 | 14.1 | -128±3 | |
| 734 | 23 08 01.3  +55 49 24 | 108.7 | -4.2 | 16.8 | -154±4 | CGCS 5811 |
| 735 | 23 38 44.3  +58 16 43 | 113.5 | -3.3 | 15.8 | -119±2 | CGCS 5904 |
| 736 | 23 46 48.8  +58 08 41 | 114.5 | -3.7 | 13.5 | -70±1 | CGCS 5932 |
| 737 | 00 51 04.9  +59 30 15 | 122.9 | -3.4 | 12.0 | -107±4 | CGCS 123 |
| 738 | 01 12 42.0  +59 44 50 | 125.6 | -3.0 | 13.4 | -80±1 | CGCS 186 |
| 805 | 01 20 00.5  +54 02 26 | 127.2 | -8.6 | 21.4 | -117±3 | CGCS 210 |

| 807 | 02 03 31.6  +55 52 15 | 133.0 | -5.6 | 24.5 | -86±4 | CGCS 6023 |
| 809 | 02 09 23.2  +54 27 51 | 134.2 | -6.7 | 20.1 | -107±4 | |

The Galactocentric distances for our stars were computed assuming a distance between the Sun and Galactic center $D_{Sun} = 7.62 \pm 0.32$ kpc ( Eisenhauer et al.) [ 12 ]. The velocities relative to the LSR were computed adopting the Solar motion from the Dahnen and Binney [ 13 ]. Finally, the rotation velocities were computed as described in section 5 of the paper [ 6 ].

*4. Results And Discussion*. Intermediate – age C stars belong to the thin disk. Figure 2 presents a plot of the mean Galactic latitude of what we call C stars, some 4400 of them (as defined by their 2MASS colors ).

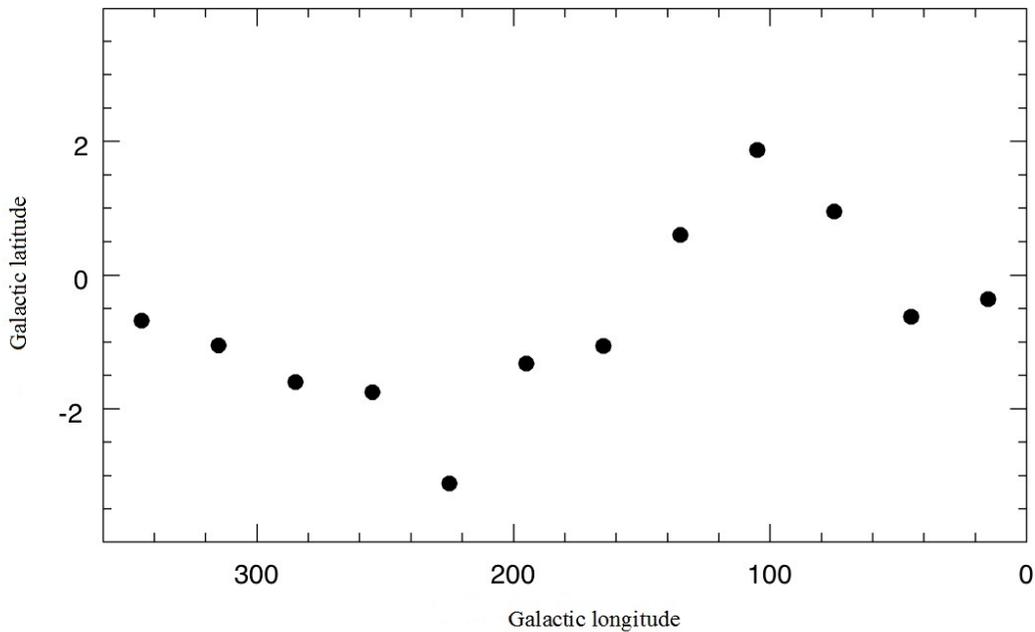

Figure 2.   Plot of the mean Galactic latitude( in degrees ) of the C stars outside the Solar circle. The means are taken over a 30º bin in longitude.

The means are taken over $30^o$ bin in longitude. We select only stars outside the Solar circle and with Galactocentric distances less than 30 kpc, to exclude extraga – lactic objects of similar NIR colors. We see a well defined warp, confirming the results by Momany et al. [ 14 ] for the thin disk. In that paper authors, using the RGB stars, detected an extended stellar population out to a distance of 24 kpc from the Galactic center and a stellar warp similar to that of Figure 2.

Our results are summarized in Figure 3 where the stars from the paper [ 6 ] are also included. The upper panel shows the circular velocities as a function of the Galactocentric distances for all C stars with $54^o < l < 150^o$. The error bar are computed taking into account the combined effects of the uncertainties in the Heliocentric distances and velocities on the derived rotation velocities. The dot with a circle corresponds to star 712. This star is likely associated with the Sagittarius dwarf spheroidal galaxy. They are separated by $35^o$ are at the same Galactocentric distance and their Heliocentric velocities differ by 10 km/s. The dashed line traces the canonical rotation velocity of 220 km/s. Having now a total 71 disk carbon stars we bin their Galactocentric distan - ces into twelve bins of six ( 5 in the first bin ). These 12 mean velocities are displayed in the lower panel. The error bars reflect the dispersion of the six velocities in each bin. The solid line marks the mean velocity of $< V > = 205$ km/s. In this panel the dashed line is a least square solution indicating a slight decline.

The new data presented here, allow us to trace the roration curve to 24 kpc, an extension of some 9 kpc to the 2007 results. The results presented in Fig. 3 show that most stars farther than 12 kpc have rotation velocities less than 200 km/s and the trend confirms that the rotation curve of the Galaxy is declined beyond the Solar circle, as suggested in paper [ 6 ].

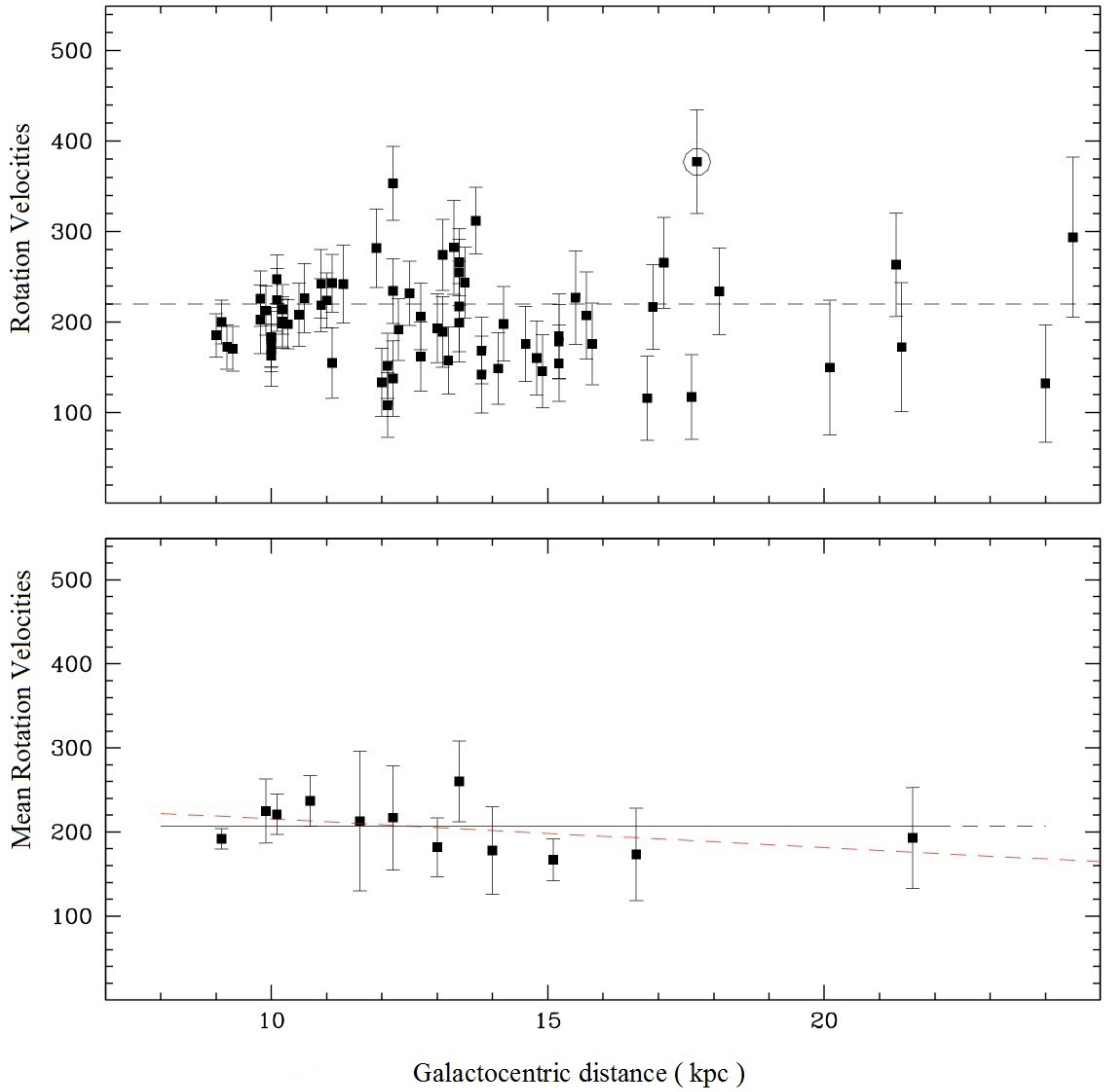

Figure 3. Upper panel displays the rotation velocity ( in km/s ) of the full set of stars. The lower panel shows the rotation curve when the stars are binned by group of six. Error bars are explained in the text.

*Acknowledgements*. This paper is funded, in parts, by the Natural Science and Engineering Research Council of Canada. The Authors would like to thank Dimitry Monin for his valuable advice that help us to achived successful runs. We are grateful to D. Bohlender for providing nights of service mode at DAO and H. Navasardyan at Asiago for the assistance during the observation. This publication makes use of data products from the Two Micron All – Sky Survey, which is a joint project of the University of Massachusets and the Infrared Processing And Analysis Center/California Institute of Technology, funded by the National Aeronautics and Space Administration and the National Science Foundation.